\documentclass{aa}  
\usepackage{graphicx}
\usepackage{txfonts}
\usepackage{natbib}
\bibpunct{(}{)}{;}{a}{}{,} 
\def\degs{\ifmmode ^{\circ}\else$^{\circ}$\fi}
\begin{document}
   \title{The changing accretion states of the intermediate polar \mbox{MU Camelopardalis}\thanks{Based on observations obtained with XMM-Newton, an ESA science mission with instruments and contributions directly funded by ESA Member States and NASA}}


   \author{A. Staude
           \and
           A. D. Schwope
           \and
           R. Schwarz
           \and
           J. Vogel
           \and
           M. Krumpe
           \and
           A. Nebot Gomez-Moran
                    }

   \offprints{A. Schwope}

   \institute{Astrophysikalisches Institut Potsdam,
                An der Sternwarte 16, D-14482 Postdam\\
              \email{astaude@aip.de, aschwope@aip.de}
             }

   \date{Received ; accepted }

  \abstract
   {}
   {We study the timing and spectral properties of the intermediate polar MU
   Camelopardalis (1RXS J062518.2+733433) to determine the accretion modes and
   the accretion geometry from multi-wavelength, multi-epoch observational data.}
   {Light curves in different observed energy ranges (optical, UV, X-ray) are
   extracted. The timescales of variability in these light curves are
   determined using Analysis of Variance. Phase-resolved X-ray spectra are
   created with respect to the most prominent detected periodicities and each
   fitted with an identical model, to quantify the differences in the fitted
   components.} 
   {The published tentative value for the spin period is
   unambiguously identified with the rotation period of the white dwarf. We
   detect a distinct soft X-ray component that can be reproduced well by a black body.
   The analysis of data obtained at different epochs demonstrates that the
   system is changing its accretion geometry from disk-dominated to a
   combination of disk- plus stream-dominated, accompanied with a 
   significant change in brightness at optical wavelengths.} 
   {}


   \maketitle
%

\section{Introduction}

\begin{figure*}[htb]
\begin{center}
\includegraphics[height=0.9\linewidth, angle=-90,clip=]{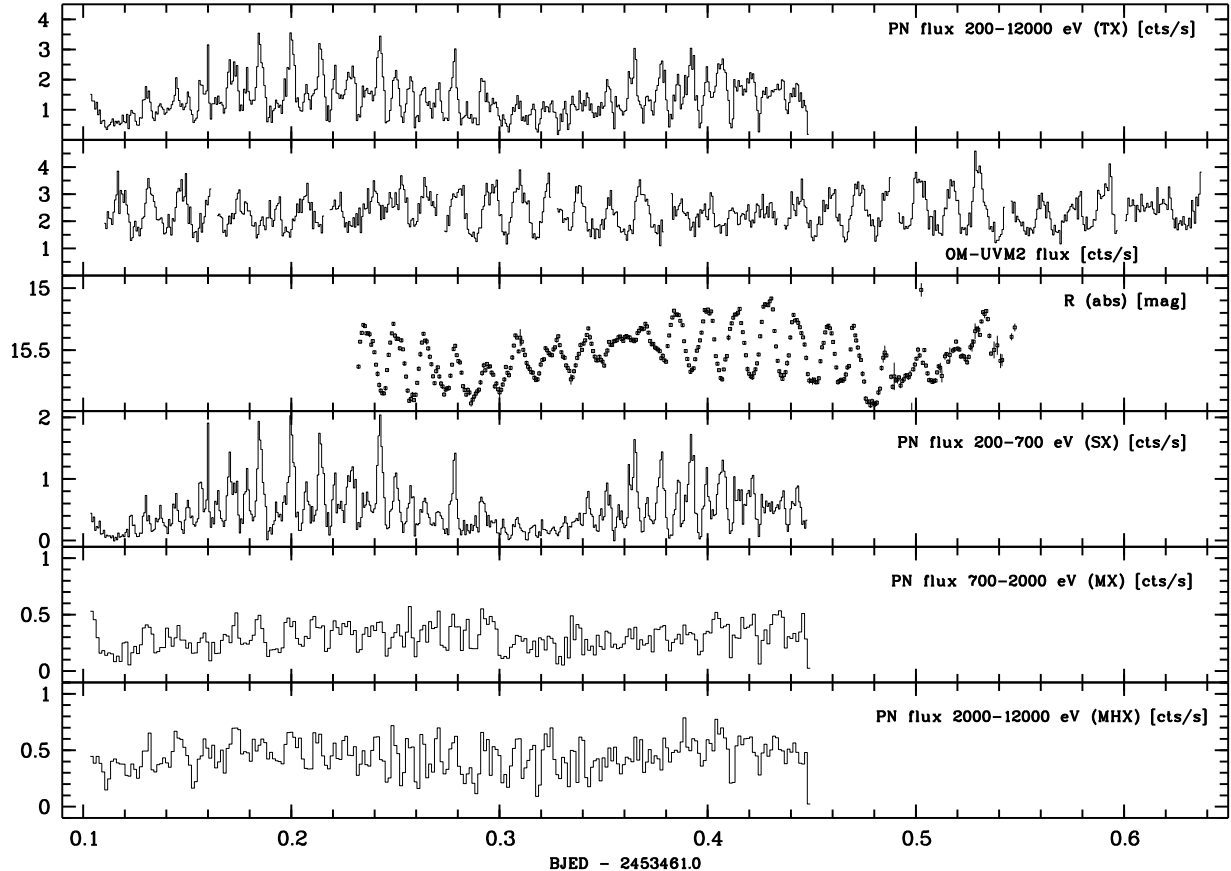}
\end{center}
\caption{\label{fig:lc2005}The light curves of MU Cam, obtained simultaneously
  on March 31, 2005, by XMM (PN and OM) and with the AIP 70cm-telescope
  (R). The data shown here are binned in time, 60s for the first, second, and
  forth panel (counted from top to bottom), 120s for the lower two. The R-band
  data are unbinned.} 
\end{figure*}

\begin{figure*}[htb]
\begin{center}
\includegraphics[height=0.9\linewidth, angle=-90,clip=]{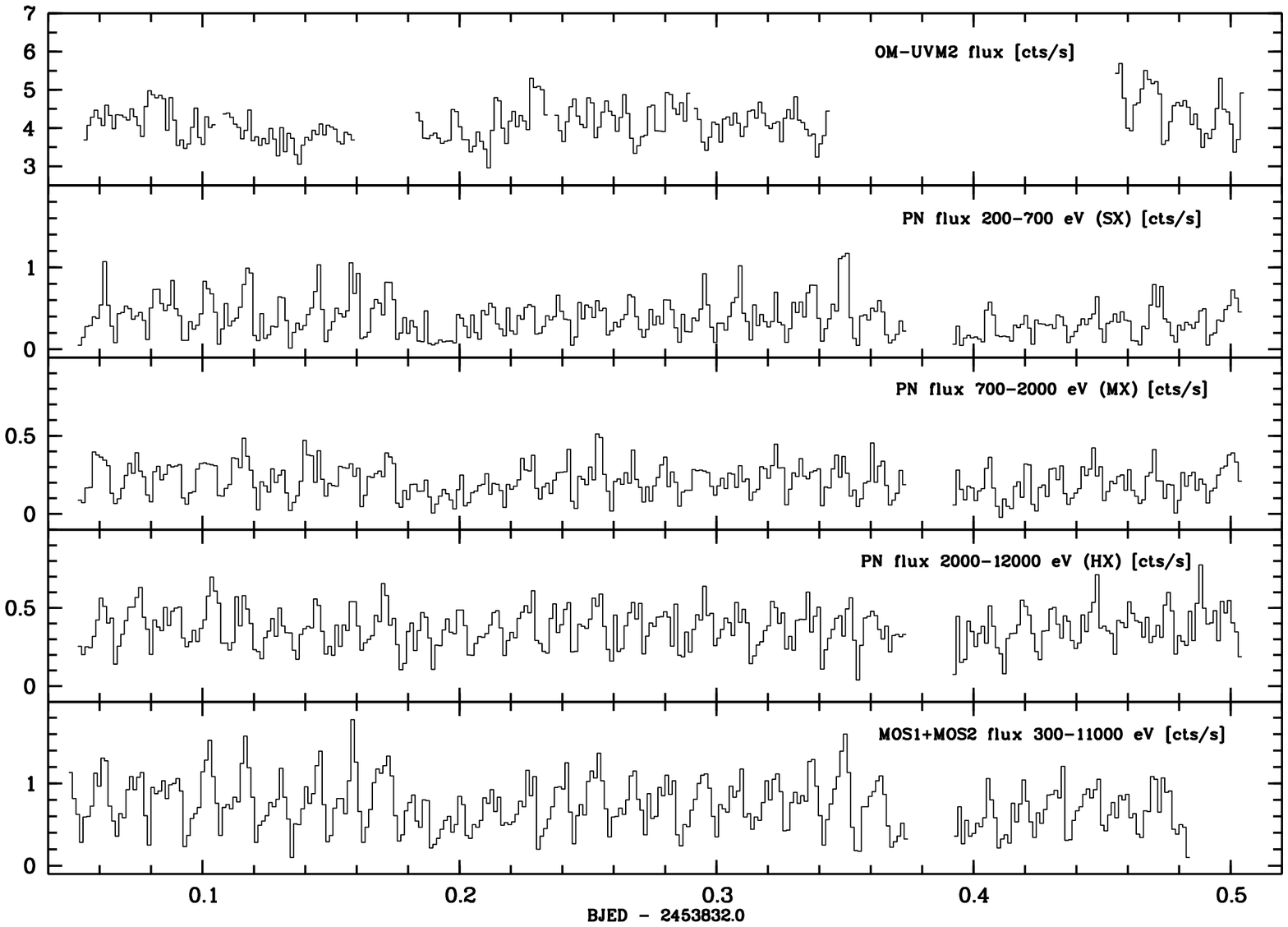}
\end{center}
\caption{\label{fig:lc2006}The light curves of MU Cam, obtained on April 6, 2006, by XMM (PN, MOS1, MOS2, and OM). The data shown here are binned to 120s.}
\end{figure*}

Intermediate Polars (IPs) are close binaries that consist of a magnetic white
dwarf and a Roche lobe-filling late-type main-sequence star. Mass loss from
the secondary star occurs via the inner Lagrangian point, and the mass is accreted by the
white dwarf. In contrast to the true Polars, the white-dwarf rotation is 
unsynchronised with the orbital motion but is more rapid due to the angular momentum
gained from the accreted matter.

In the case of a high accretion rate or a weak magnetic field, the
magnetosphere will be smaller than the circularisation radius and an accretion
disk will form. The inner disk is truncated by the magnetic field of the white
dwarf, which channels the matter via accretion curtains to the magnetic
poles. Accretion arcs are formed along the footpoints of accreting field lines.
Along these arcs, the accretion rate is
assumed to vary.  The heated plasma in the accretion zone is a
prominent site of hard X-ray radiation, which due to the lighthouse effect is
modulated on the spin period, $\omega$. Because of irradiation, the inner-disk
rim may emit radiation that is modulated on this period.

For stronger magnetic fields, some or all of the material from the infalling stream
might be directly captured by the field without going through an intermediary
accretion disk. In the case of this stream-fed accretion, material transferred to
the white dwarf retains the orbital motion of the
secondary star, producing periodic variability of the beat frequency
$\omega$-$\Omega$, or other sidebands with the orbital frequency $\Omega$.

Most IPs accrete predominantly by means of an accretion disk, V2400 Oph being
a rare example of a system with pure stream-fed accretion
\citep{hellierbeardmore2002}. A number of IPs show evidence for accretion by
means of both mechanisms simultaneously: via accretion curtains beginning at the
inner disk rim, and via so-called disk-overflow when a part of the ballistic
stream bypasses the accretion disk outside the orbital plane. In some
systems repeated changes between the different accretion states were observed
(TX\,Col, \citealt{norton1997}; FO\,Aqr, \citealt{evans2004a}). 
Principal
evidence for stream-overflow in IPs has been based on indirect means, i.e.~the
detection of certain sideband frequencies in their power spectra, which is
not always unambiguous. For instance, sideband frequencies may also be produced
by spin-pulsed light reprocessed in orbitally-fixed structures (a problem
mainly in the optical and UV); alternatively, 
amplitude modulation of the orbital period
(e.g.~due to X-ray absorption) may introduce spurious side-bands
\citep{warner1986}.

An article by \citet{patterson1994} introduces IPs in general, whereas
\citet{hellier2002} discusses the mechanisms of stream-fed accretion.

The X-ray source MU Camelopardalis (the new name according to
\citet{2006IBVS.5721....1K} for the source formerly known as
1RXSJ062518.2+733433) was identified as an intermediate polar by
\citet{sab2003} and \citet{staude2003a}. Optical photometry showed strong
variability on mainly two timescales, which were tentatively identified with
the orbital period of the system and the spin period of the white dwarf,
\mbox{$P_{\rm orb} = 16987(23)$\,s$~= 0.19661(27)$\,d} and \mbox{$P_{\rm spin}
  = 1187.246(4)$\,s$~= 0.01374127(5)$\,d}, respectively
\citep{staude2003a}. 
It remained an open question whether the measured spin period provided the
white-dwarf rotation, if it was twice the true rotation value, or if it was a
side-band period with the orbital variation. 

MU Cam is one of a handful of IPs that show a prominent soft X-ray
component additional to the hard component \citep{staude2003a}. An
XMM-Newton observation was proposed to identify undoubtedly the spin period of
the white, and to determine the origin of the soft X-ray radiation.

We report on X-ray and UV observations of MU Cam with XMM-Newton on both March 31,
2005, and April 6, 2006, and coordinated ground-based optical photometry. 


\section{Observations}

A short summary of all observations analyzed in this work is provided in Table \ref{tab:obslog}.

\begin{table}
\caption{\label{tab:obslog}The observations analyzed in this work}
\begin{center}
\begin{tabular}{l l r}
\noalign{\smallskip} \hline
\hline
\noalign{\smallskip} 
Date (dd/mm/yyyy)  & Telescope/Instrument     & Duration\\
\noalign{\smallskip} \hline \noalign{\smallskip}
28/03/2005         & AIP 70cm / R               & 5.0\,h\\
30/03/2005         & AIP 70cm / R               & 8.6\,h\\
20/04/2005         & AIP 70cm / R               & 4.7\,h\\
31/05/2005         & AIP 70cm / R               & 7.6\,h\\
31/05/2005         & XMM / PN+MOS+OM            & 12.5\,h\\
\noalign{\smallskip} \hline \noalign{\smallskip}
06/04/2006         & XMM / PN+MOS+OM            & 10.6\,h\\
08/05/2006         & AIP 70cm / R               & 4.4\,h\\
\noalign{\smallskip} \hline
\end{tabular}
\end{center}
\end{table}

\subsection{Data from XMM-Newton}

The X-ray counting EPIC instruments were operated in small-window mode
with the thin filter. The EPIC-RGS devices were used in the standard
configuration. The optical monitor (OM) was kept in fast mode with filter
UVM2.

On March 31, 2005 (revolution 972), an observation with XMM-Newton (observation
ID: 0207160101), of a scheduled duration of 49\,ksec, was performed. Due to the
high background, both EPIC-MOS detectors were switched off after 13.6\,ks,
while the EPIC-PN observed for 30\,ksec. The OM was in operation for
45\,ksec. Ground-based observations were performed synchronously in the
R-band. The X-ray observations corresponded to 1.75 binary orbits and 25 white-dwarf
spin cycles, in which the EPIC-PN collected $\sim$37,000 source counts. A
preliminary analysis was presented by \citet{staude2006}. The observed
mean source count rates were 1.54\,s$^{-1}$ in the PN,
0.38\,s$^{-1}$ and 0.43\,s$^{-1}$, in the MOS1 and MOS2, respectively; 2.3\,s$^{-1}$ in the OM/UVM2;
and 0.031\,s$^{-1}$ and 0.040\,s$^{-1}$, in the first order of RGS1 and RGS2,
respectively.

In revolution 1158 (on April 6, 2006, observation ID: 0306550101), XMM-Newton
observed the target again for a scheduled $\sim$39\,ksec. All instruments were
used for the complete observation duration. Due to a telemetry gap, half an
hour of data is missing for all instruments. Additionally, two blocks of
4400\,s exposure in the OM were also lost. During this observation, the mean
source count rates were measured to be 1.17\,s$^{-1}$ in the PN; 0.30\,s$^{-1}$
and 0.34\,s$^{-1}$ in MOS1 and MOS2, respectively, 4.2\,s$^{-1}$ in the OM/UVM2; and
0.022\,s$^{-1}$ and 0.029\,s$^{-1}$ in the first order of RGS1 and RGS2, respectively.

The data reduction was completed using {\sc XMM-SAS} version 6.5; in
particular the tasks
{\sc epchain}, {\sc emchain}, and {\sc rgsproc} were adopted to process the X-ray
data. Source and background photons were extracted in adjoining circles for
the PN and for the MOS data from 2006, while the background in the MOS
observation of 2005 was extracted from an annulus around the source
circle. Events with detection patterns of up to quadruples were selected.

Background-subtracted light curves in different X-ray energy bands -- created
with {\sc evselect} -- of the PN data are shown in Figs.~\ref{fig:lc2005} and
\ref{fig:lc2006} with the OM UV-data and  ground-based R-band data acquired
contemporaneously.
The energy bands were selected after inspection of the
shape of the averaged spectra (Fig.~\ref{fig:EPIC_spec}), which showed a soft
component below 0.7\,keV, and a minimum in the harder energy range
between 2 and 3\,keV. We therefore define the following PN energy bands: \mbox{{\it
    SX}: $0.3-0.7$\,keV}, {\it MX}: $0.7-2.0$\,keV, {\it HX}:
$2.0-12.0$\,keV, {\it MHX}: $0.7-12.0$\,keV, and {\it TX}:
$0.3-12.0$\,keV.

EPIC source spectra were created with the task {\sc especget}. RMF files for
the data from the RGSs were created with {\sc rgsrmfgen}. In further
analysis, all spectra were grouped with {\sc grppha} to contain at least
20 counts per bin.

The OM light curves were extracted using the task {\sc omfchain}.

\subsection{Optical photometry}

Photometric observations in the optical wavelength range were performed with
the 70cm telescope of the AIP in Potsdam. Differential photometry for this data
was completed using DoPhot. The apparent R-band magnitudes were calculated from the
brightness of the comparison star as in \citet{staude2003a}.

Simultaneously to the observations with XMM-Newton on March 31, 2005, 7.6\,h of
optical data were acquired with exposure times of 60\,s (see
Fig.~\ref{fig:lc2005}). For the period analysis, further data taken on March
28 and 30, and on April 20, 2005 were included.

Compared to the R-band data presented by \citet{staude2003a}, the object was almost 
one magnitude fainter in 2005. In 2003, the optical short-term variability
occurred on the period that we would now identify with the true white-dwarf
spin (see Sect.~\ref{period_analysis}), whereas in 2005 it occurred on
$\omega$-$\Omega$. 

\begin{figure}[htb]
\includegraphics[height=\linewidth, angle=-90,clip]{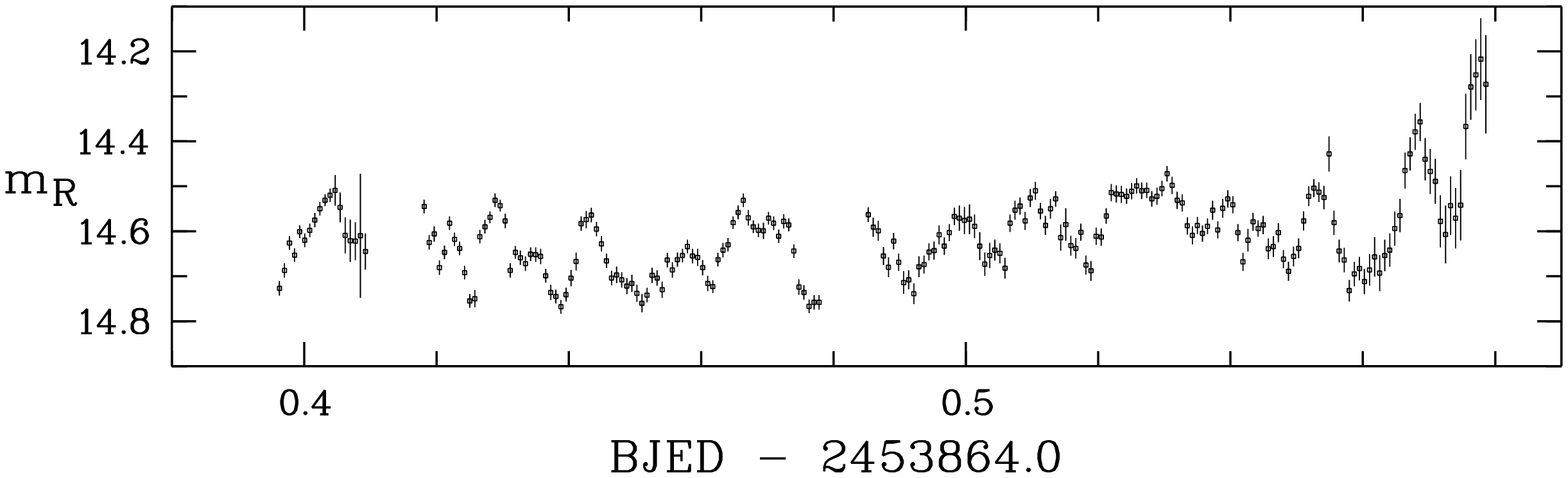}
\medskip\\
\includegraphics[height=\linewidth, angle=-90,clip]{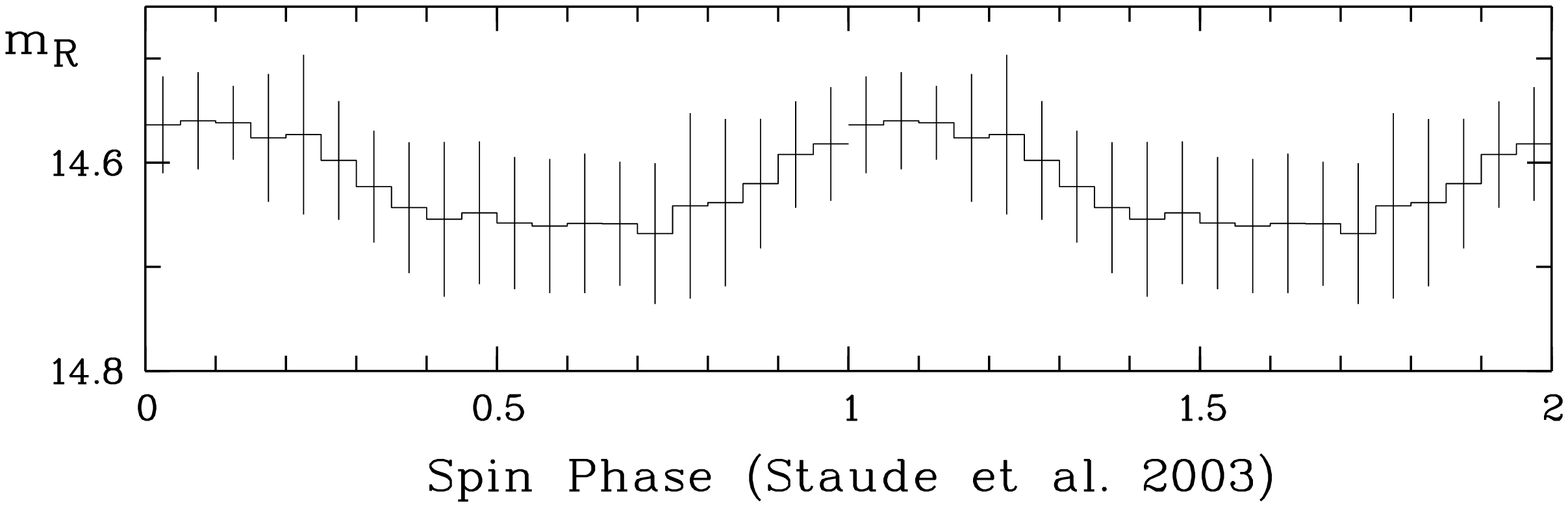}
\caption{\label{fig:rx0625_lc_08.05.2006} {\it top}: $R$-band light curve from
  May 8, 2006, {\it bottom}: The above data with BJED$\le$2453864.56, binned
  into 20 bins on the spin period calculated with the ephemeris given by
  \citet{staude2003a}. The error bars mark the standard deviation of the
  values entering each bin.} 
\end{figure}

There is no optical photometry simultaneous to the XMM-Newton observation of
2006. Nevertheless, the star was observed for 4.4\,h on May 08, 2006, using
the AIP 70cm-telescope with 60\,s exposure time
(see Fig.~\ref{fig:rx0625_lc_08.05.2006}).

As found for data acquired in 2003, the system was measured to be one
magnitude brighter than in 2005 and its variability to be dominated by the
white-dwarf spin period (see Sect.~\ref{period_analysis}).

\section{Analysis of the averaged X-ray spectra}
\label{chap:spec_anal}

\begin{figure}[htb]
\includegraphics[width=\linewidth,angle=0,clip=]{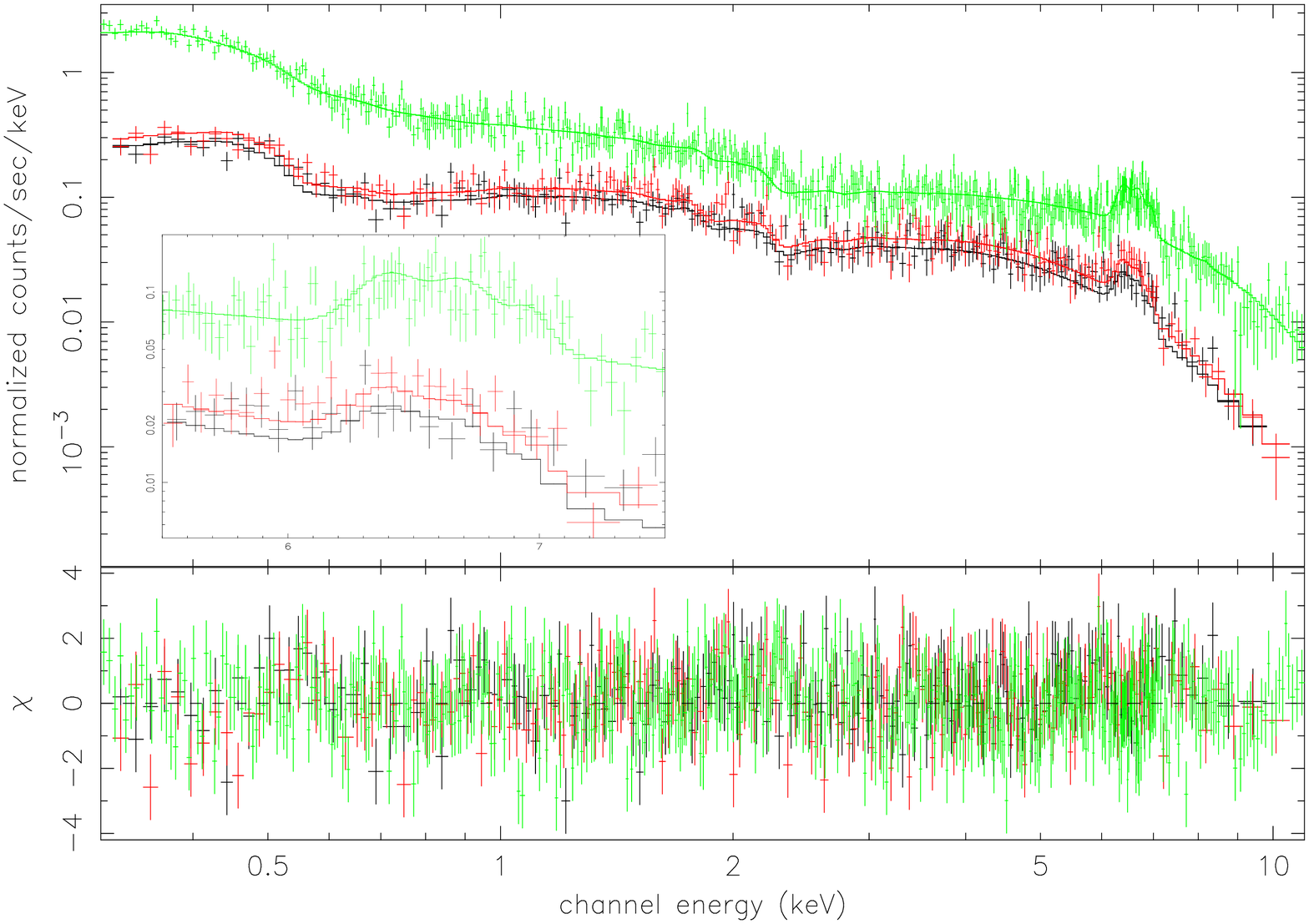}
\includegraphics[width=\linewidth,angle=0,clip=]{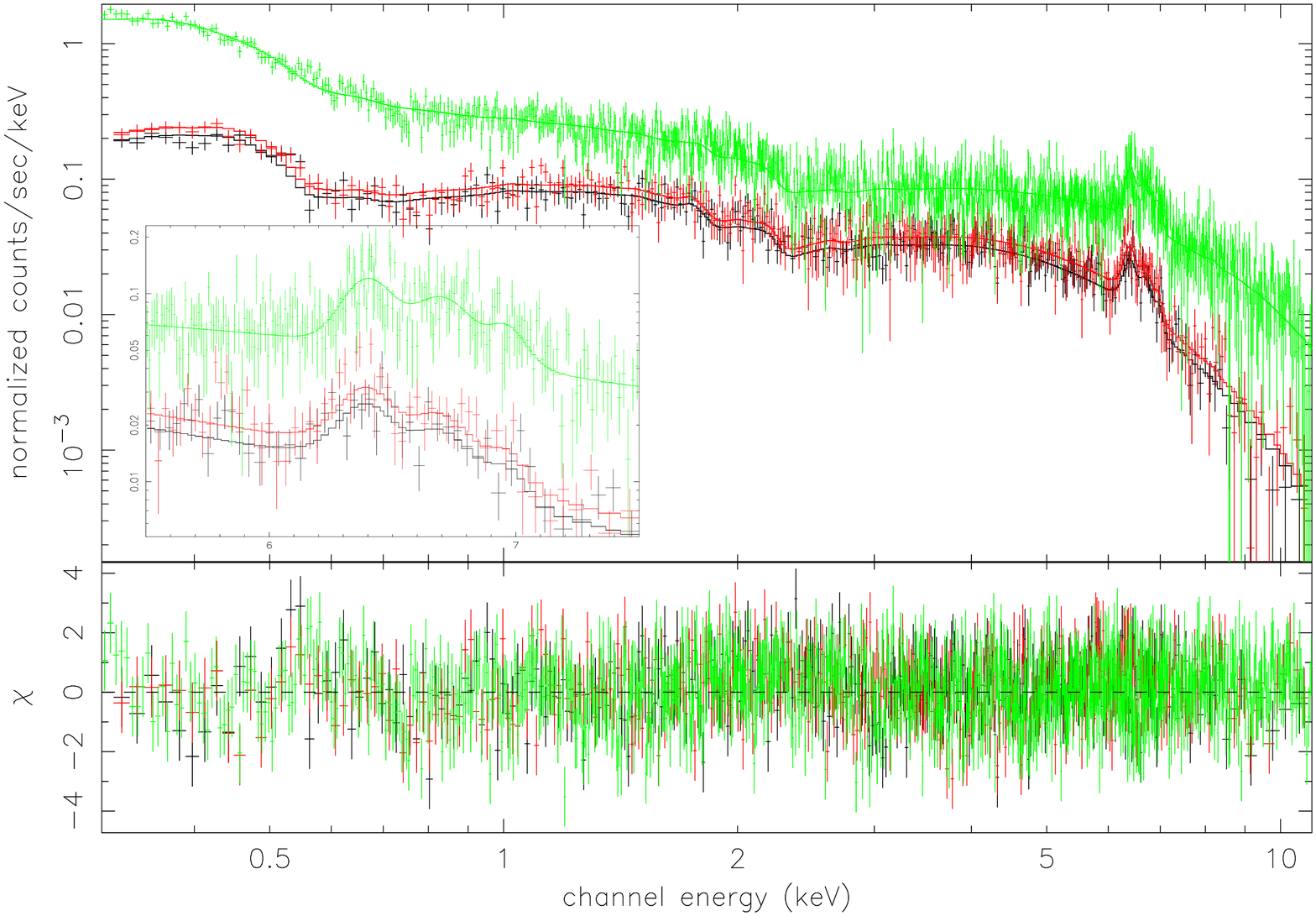}
\caption{\label{fig:EPIC_spec}The EPIC X-ray spectra from 2005 ({\it top}) and
  2006 ({\it bottom})(green PN, red+black: MOS1/2) with overplotted fit and
  residuals (Parameters in Tab.~\ref{tab:pars_spec}). The insets show an
  enlarged view of the iron-K region.} 
\end{figure}

\begin{table}
\caption{\label{tab:pars_spec}The main parameters of the spectral model for
  the average XMM-Newton spectra from 2005 and 2006. The numbers in brackets are the
  errors and have to be multiplied by the same power of ten as to that for the
  parameters. `n.h.p.' is the `null hypothesis probability'.} 
\begin{center}
\begin{tabular}{l l l l}
\noalign{\smallskip} \hline
\hline
\noalign{\smallskip}
                &                          & 2005              & 2006 \\
\noalign{\smallskip} \hline \noalign{\smallskip}
wabs            &nH [10$^{22}$\,cm$^{-2}$] &4.1E-02 (0.8)   & 4.7E-02 (0.6)\\
\noalign{\smallskip} \hline \noalign{\smallskip}
pcfabs          &nH [10$^{22}$\,cm$^{-2}$] &7.9 (0.6)   & 9.5 (0.5)\\
                &CvrFract                  &0.61 (0.02) & 0.67 (0.01)\\
\noalign{\smallskip} \hline \noalign{\smallskip}
mekal           &kT [keV]                  &79.9 [35]$^{1}$ & 79.9 [35]$^{1}$ \\
                &norm        &7.3E-03 (0.4) & 6.5E-03 (0.3)\\
\noalign{\smallskip} \hline \noalign{\smallskip}
bbody           &kT [keV]         &5.9E-02 (0.07) & 5.4E-02 (0.05)\\
                &norm        &1.9E-04 (0.9) & 2.2E-04 (0.9)\\
\noalign{\smallskip} \hline \noalign{\smallskip}
gaussian  &LineE [keV]      &6.4 (-- $^{2}$) & 6.4 (-- $^{2}$)\\
                &Sigma [keV]      &0.1 (-- $^{2}$) & 0.07 (0.02)\\
                &norm       &3.1E-05 (0.7) & 2.9E-05 (0.6)\\
\noalign{\smallskip} \hline \noalign{\smallskip}
gaussian  &LineE [keV]      &6.7 (-- $^{2}$) & 6.7 (-- $^{2}$)\\
                &Sigma [keV]      &0.1 (-- $^{2}$) & 0.09 (0.03)\\
                &norm       &3.1E-05 (0.7) & 2.5E-05 (0.7)\\
\noalign{\smallskip} \hline \noalign{\smallskip}
$\chi^{2}_{\nu}$ (d.o.f) &              & 1.034659 (948) & 1.046327 (1754)\\
\noalign{\smallskip} \hline \noalign{\smallskip}
n.h.p.&                & 2.2E-01 & 8.7E-02\\
\noalign{\smallskip} \hline
\end{tabular}
\end{center}
{\footnotesize 
$^{1}$: The temperature of the {\it mekal} component ran against its upper
limit. The value in brackets is the plasma temperature which is compliant with
both the XMM and {\it INTEGRAL} spectrum (see Fig.~\ref{fig:MUCam_SED} and
chapter \ref{chap:spec_anal}).\\ 
$^{2}$: The parameters were fixed to these values.
}
\end{table}

The EPIC-PN and MOS1/2 spectra for both observations are shown in
Fig.~\ref{fig:EPIC_spec}. Both spectra appear similar and have flux in all
energy bins in the interval from 0.3 to 11\,keV. In the energy range below
0.6\,keV, a distinct soft component is observed, which confirms the classification
of MU Cam as a soft IP. Around 6.5\,keV, contributions from \ion{Fe}{} emission
lines are detected. The spectral analysis of our
data was performed with Xspec 11.3.0.

\citet{staude2006} fitted the spectrum from 2005 with a model consisting of a
hot emitting gas ({\it mekal}), a black body component ({\it bbody}), and a
Gaussian representing iron-K emission lines ({\it gaussian}) absorbed by a
photoelectric absorber ({\it wabs}), and an additional hot-gas model absorbed
by a photoelectric absorber of different column density, i.e.~{\it
  wabs$_{1}$(mekal$_{1}$)+wabs$_{2}$(mekal$_{2}$+bbody+gaussian)}. Since the
hydrogen column density, $n_{\rm H}$, of {\it wabs$_{1}$} was 250 times
higher than that of {\it wabs$_{2}$} ($n_{\rm
  H}$=2.7$\times$10$^{22}$\,cm$^{-2}$), the model consisted of components
absorbed by a hydrogen column comparable to the interstellar absorption, and
an additional highly-absorbed {\it mekal}.

The data from 2005 and 2006 were fitted in the range from 0.3 to 11\,keV with
Xspec using an identical multi-component model for both data sets. This model
consisted of an emitting hot-gas component ({\it mekal}), a black body ({\it
  bbody}), and two Gaussian emission lines representing the iron-K lines at
6.4\,keV and 6.7\,keV ({\it gaussian}), which were all absorbed by a
partially-covering photoelectric absorber ({\it pcfabs}), and a simple absorber 
representing the interstellar medium ({\it wabs}), i.e.~\mbox{{\it
    wabs(pcfabs(mekal + bbody + gaussian + gaussian))}}. 
The model fitted the data as well as the previous model, although it required
just one partially-absorbed hot-gas component, instead of two of different
homogeneous absorption. 
The model can be understood according to the standard picture of
IPs: The {\it bbody} represents the soft X-ray emission below 0.7\,keV, either
thermal emission or reprocessed hard X-rays, the {\it mekal} component
describes the hard component emitted by the hot post-shock material above the
white dwarf's surface, and the {\it gaussian} emission line at 6.4\,keV (the
\ion{Fe}{} fluorescence line) are tracers of reprocessed X-rays. The {\it pcfabs}
models absorption inside the system by the accretion curtain/stream. It is
mainly responsible for the absorption of soft photons in the {\it mekal}
component, which explains the relative weakness of flux in the range between 0.7
and 7\,keV with respect to flux at the harder end of the spectrum.

Values of the free-fit parameters were determined by $\chi^{2}$-minimisation
and are listed in Table \ref{tab:pars_spec}. The fits were successful in both
cases with a reduces $\chi_\nu^{2} \sim 1$. Adding further components such as
another {\it  mekal} or a black body, improved the fit only marginally. A
third Gaussian was not required to model the \ion{Fe}{}-line at 6.96 keV, because this
line was described well already by the {\it mekal} model.
Our fit was not influenced significantly by allowing the metal abundance of
the {\it mekal} to vary as a free parameter; in this case, the {\it mekal}, in fact,
hardly changed from its initial value.
The lines of iron
were resolved in data acquired in 2006 and their widths were therefore allowed
to vary, whereas they were unresolved
in the 2005 data (see insets in Fig.~\ref{fig:EPIC_spec}). In the data from 2005, there is line emission at the
positions of all three iron-K lines, although an insufficient number of counts
do not enable the lines to be resolved. We therefore fixed their line widths to
0.1\,keV in the fitting, assuming that they were as broad as in 2006.

The parameters of the partial absorber in our fits are comparable to the
values obtained for the other soft IPs observed with XMM-Newton, V405 Aur
($n_{\rm H}$=6.1$\times$10$^{22}$\,cm$^{-2}$, $CvrFract$=0.52,
\citealt{evans2004b}), PQ Gem ($n_{\rm H}$=11.1$\times$10$^{22}$\,cm$^{-2}$,
$CvrFract$=0.45, \citealt{evans2006}), UU Col ($n_{\rm
  H}$=10$\times$10$^{22}$\,cm$^{-2}$, $CvrFract$=0.51,
\citealt{demartino2006}), and NY Lup ($n_{\rm
  H}$=9.7$\times$10$^{22}$\,cm$^{-2}$, $CvrFract$=0.47,
\citealt{haberl2002}).

A problem, however, with our fits is that the fitted temperatures from the 
{\it mekal} models are at the upper limit for the model available in Xspec. 
Reducing the
temperature to lower values, produces fits in which the {\it pcfabs\_nH} is
reduced but its covering-fraction increased; oervall have we a higher
$\chi_\nu^{2}$. \citet{haberl2002} report that fitting a multi-temperature
distribution of {\it mekal} models with temperatures following a power law to
the data of NY Lup, leads to an unconstrained maximum temperature, i.e.~a
similar problem. 

\begin{figure}[htb]
\includegraphics[height=\linewidth,angle=-90,clip=]{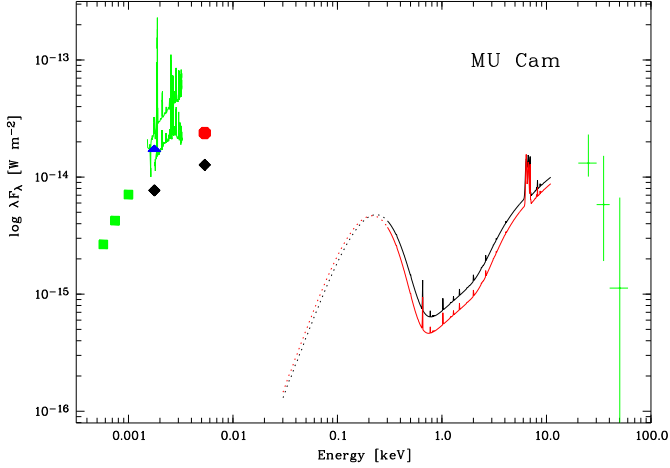}
\caption{\label{fig:MUCam_SED} The spectral energy distribution of MU Cam.
 Shown are the data synchronously obtained in 2005 (rhombs and X-ray model
 spectrum in {\it black}), the data from 2006 (XMM-Newton: circle and X-ray
 model spectrum in {\it red}, {\it R}-band flux from the 70cm telescope: {\it
   triangle}), and, additionally, 2MASS {\it J, H, and K} fluxes ({\it
   squares}) plus the optical spectra from \citet{sab2003}~({\it lower}) and
 \citet{wei1999}~({\it upper}). 
Additionally, the three data points from the {\it INTEGRAL}/IBIS spectrum
\citep{barlow2006} are plotted with their corresponding error bars. 
 The X-ray models are plotted without the interstellar absorption
 component. To illustrate the shape of the black body component in the X-rays,
 the extrapolation to lower energies outside the fitted range is plotted with
 dots.  
}
\end{figure}

MU Cam was detected to be an {\it INTEGRAL}/IBIS source. \citet{barlow2006}
measured a bremsstrahlung temperature of 8.1(4.7)\,keV. Judging from the
XMM-data, this value is clearly too low; we therefore analysed both data sets
together. The {\it INTEGRAL} spectrum discussed in \citet{barlow2006} was
kindly provided by the authors and is included in Fig.~\ref{fig:MUCam_SED}.
Fitting both data sets simultaneously with Xspec produced no satisfying results
because of the large differences in the number of data points and their
errors.  However, by fixing the temperature of the {\it mekal}-component to a
number of different values and fitting the remaining parameters, good fits to the
  XMM-data ($\chi_\nu^{2}<1.1$) and a convincing agreement with the {\it INTEGRAL}
  data points (the model was located within the error-boxes of the three data points)
  were achieved for $kT = 35 \pm 10$\,keV.

Utilizing instruments capable of observing a wider range of X-ray energies
(e.g.~RXTE), the plasma temperatures in IPs can be determined more precisely
\citep{suleimanov2005}. Compared to the temperatures measured for other IPs,
the value of $kT$=35\,keV is unusually high. Typical values for IPs are
10-20\,keV. Since this temperature must be regarded as an average of the
entire temperature range, and since the reflection component present in the
star redistributes photons to higher energies, 
the post-shock temperature remains only weakly constrained. 

The EPIC-spectra and the parameter values of the fits to the two data sets of
2005 and 2006 appear to be quite similar, which is surprising when differences,
particularly in the soft X-ray light curves are considered. 

The spectra from RGS1 and RGS2 of both epochs were grouped to contain at least
20 counts. The fit to the integrated EPIC spectra reproduces the data well. No
significant lines were detected in the wavelength range considered, and we
therefore disregard the RGS data in further analysis. 

The overall spectral energy distribution of the system is shown in
Fig.~\ref{fig:MUCam_SED}, where the available measurements are plotted with
colours that differentiate between data obtained simultaneously (black and
red), 
acquired in an
optically-bright state as assessed from the R-magnitude (blue), and of unknown
state (green).

During an epoch of high optical brightness, the X-ray flux is lower than
when the system is optically faint. It is evident that the black body
component, which can be seen in the X-rays, contributes only negligibly to the optical/UV
flux. The shape of the continuum in the optical/UV wavelength range resembles
that of black body with a much lower temperature and can be attributed to the
accretion disk. 

\section{Variability analysis}
\label{period_analysis}

\begin{figure*}[htb]
\mbox{
\includegraphics[height=0.5\linewidth, angle=-90,clip]{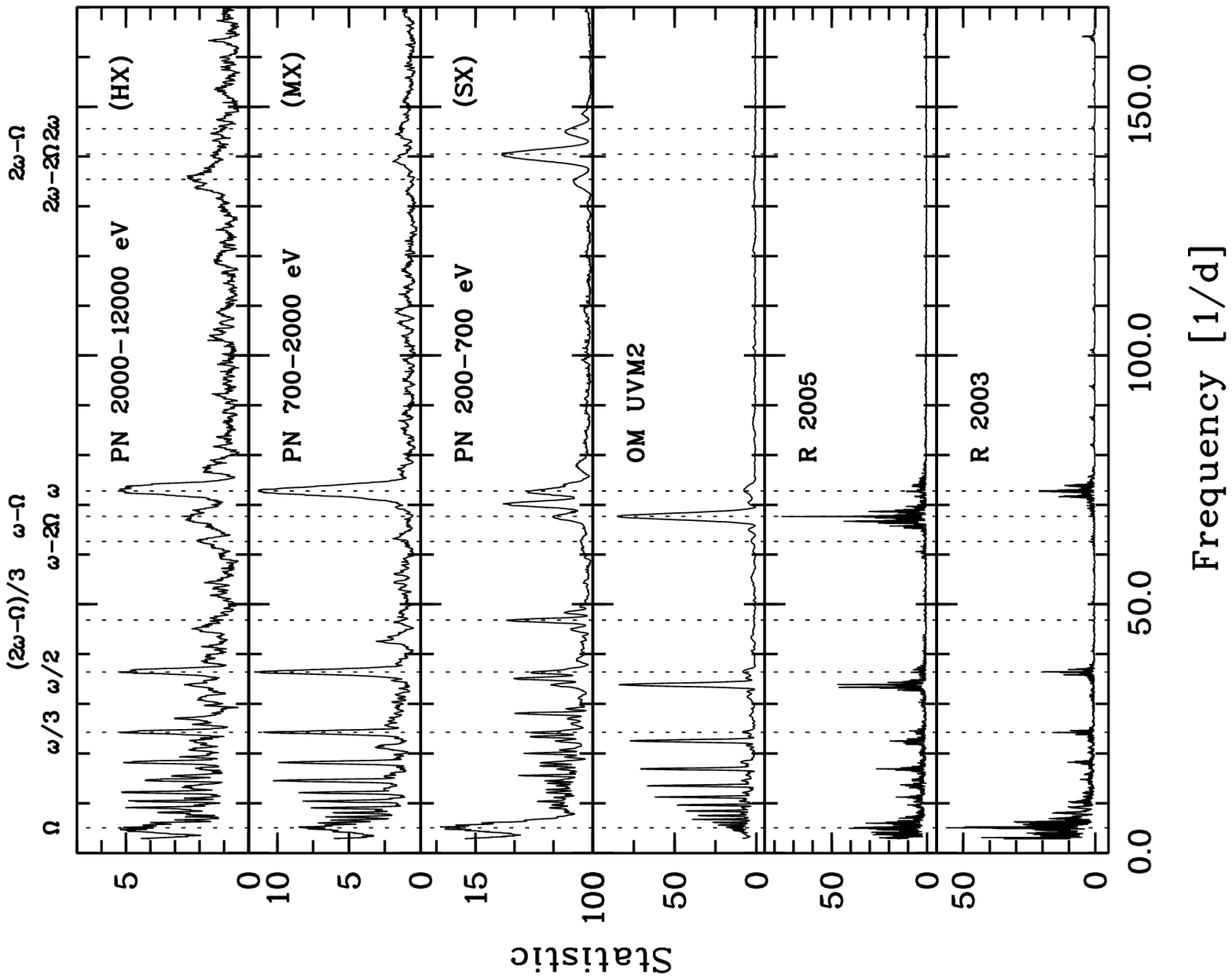}
\hfill
\includegraphics[height=0.5\linewidth,angle=-90,clip]{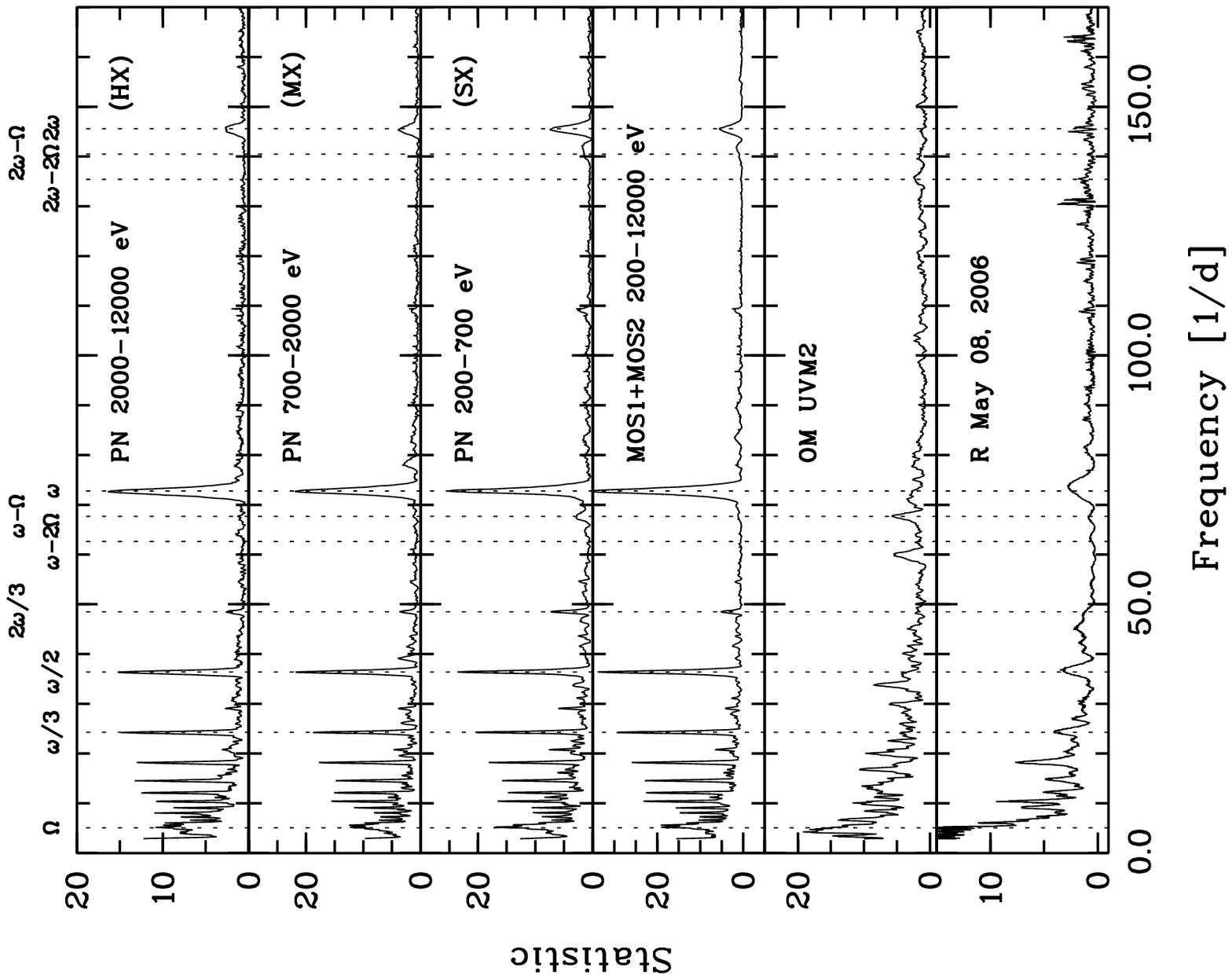}
}
\caption{\label{fig:aov}AOV-periodograms of {\it left}: The data from 2005
  (Fig.~\ref{fig:lc2005}) -- PN counts in different energy ranges, OM with
  UVM2, and $R$-band photometry -- for comparison, the AOV analysis of the
  $R$-band data from 2003 \citep{staude2003a} is added; {\it right}: The data
  from 2006 (Fig.~\ref{fig:lc2006}) in different energy ranges and
  instruments. The orbital ($\Omega$) and spin ($\omega$) frequencies, their
  sidebands, and (sub-)harmonics are marked with dotted lines.} 
\end{figure*}

The light curves derived from the observations during the two epochs are shown
in Figs.~\ref{fig:lc2005} and \ref{fig:lc2006}, the corresponding periodograms
in Fig.~\ref{fig:aov}. The analysis is supported by additional optical
photometric data that are  listed in Table \ref{tab:obslog}. In spite of
differences between data for different epochs, the periodograms provide a
clear measurement of the basic periods of the system.

\subsection{The basic frequencies}

\citet{sab2003} and \citet{staude2003a} gave a prelimininary identification of
the system orbital and spin periods from optical photometric
data. Optical spectroscopy was used by \citet{sab2003} to identify
the orbital frequency, $\Omega$. Using the optical data alone, it was impossible
to identify unequivocally the white dwarf spin frequency, $\omega$.

Figure~\ref{fig:aov} shows the results of a period analysis using the method of
analysis-of-variance (AOV, \citealt{schwarzenberg-czerny1989}) in the
different energy bands, for data from the two XMM campaigns. We indicate the orbital
frequency $\Omega$, the tentative spin frequency $\omega$, the side-bands of
both these frequencies, and the (sub)harmonics. AOV tends to create artifacts in the
form of sub-harmonics; this is in contrast to the Fourier-transform, which creates
harmonics. The main peaks in the X-ray bands {\it HX} and {\it MX} are in
both datasets at the assumed $\omega$. In the {\it SX} band, the pattern is
more complex in 2005 than in 2006, showing variability not only at $\omega$
but also at $\Omega$ and $2\omega$-$\Omega$. The OM-data show variability in
the so-called beat-frequency $\omega$-$\Omega$, at least in data from
2005. In the optical {\it R}-band, the peak is located at either $\omega$
(2003 and 2006) or $\omega$-$\Omega$ (2005).

A consistent explanation for the presence of all peaks in the periodograms can be given
when the tentative value of $\omega$ is assumed to measure the white
dwarf spin.

The X-ray periodograms from 2006, which show variability power only at $\Omega$,
$\omega$, and $2\omega$, could be produced by accretion from
a truncated disk to two opposite accretion regions via shock-heated curtains
\citep{wynnking1992}; reprocession on the white dwarf surface would then 
produce the soft X-rays ({\it SX}). In 2005, the harder X-rays ({\it
  HX} and {\it MX}) were created by accretion along curtains, indicated by
variability in mainly $\Omega$ and $\omega$. The soft X-rays ({\it SX}) have
variability power also in the frequency $2\omega$-$\Omega$ and some signal at
$\omega$-$\Omega$. This can be explained by their origin in reprocession on
the white dwarf as well as by accretion directly from the accretion stream via
magnetic field lines, which would produce modulations in superpositions of
$\Omega$ and $\omega$ \citep{wynnking1992}. 

In 2005, the UV data shows
variation mainly at $\omega$-$\Omega$, the periodogram has weak power only at
$\Omega$ and $\omega$. The {\it R}-band data has a similar variability pattern with
stronger variation in $\Omega$. Changes in the dominating frequency from
$\omega$ to $\omega$-$\Omega$, especially at the lower energies, can be
described by changing accretion modes within the current models of IPs.

We could imagine assuming the tentative $\omega$ to be the true 
$\omega$-$\Omega$. This would be consistent with diskless accretion along
magnetic field lines. Certain problems arise in this approach, for example 
the need to explain the absence of any signature of variation with frequency 
$\omega$. Since the visibility of the accretion regions is modulated by this
frequency, it is expected to be observed in hard X-rays for which absorption
is probably incomplete. In the $R$-band and UV data from 2005, one would have 
to explain the strong signal at $\omega$-2$\Omega$ (the tentative
$\omega$-$\Omega$), a small contribution to
$\omega$-$\Omega$, and no trace of $\omega$. 

\citet{warner1986}
explained an $\omega$-2$\Omega$ varibility to be an orbital modulation of the
$\omega$-$\Omega$ component. It is therefore unlikely that the latter has a far
smaller amplitude. 
We would also have to replace the convenient explanation of $2\omega$-$\Omega$
proposed by \citet{wynnking1992}, by one for $2\omega$-$2\Omega$; furthermore,
for the 2006 data, we would have to consider $2\omega$-$\Omega$ without a
signature for the $2\omega$ periodicity.

To summarize, it is unlikely that this second interpretation reproduces our
observations of MU Cam. We therefore claim that we have found the spin period
predicted by \citet{staude2003a} to be the true rotation of the white dwarf.

Aside from their common properties, the data from the two epochs show a number
of differences.

In the light curves from 2005 (Fig.~\ref{fig:lc2005}), the soft X-rays ({\it
  SX}) and the R-band data show a strong dependence of variability amplitude
on the orbital phase. In the soft X-ray light curve, two peaks per
spin-cycle of various heights are directly visible. They are probably the
origin of the frequency $2\omega$-$\Omega$.

The most striking differences between the light curves obtained on April 06,
2006 (Fig.~\ref{fig:lc2006}), and those from 2005 (Fig.~\ref{fig:lc2005}),
that there is a 
reduction of the {\it SX} countrate and the amplitude of the short-term
variation does not depend obviously on the orbital period. The
orbital period $\Omega$ is detected in the periodograms of all the X-ray
bands, although there is no obvious variation of this timescale in the light
curves (see Fig.~\ref{fig:data_orbfolded}).

The $R$-band data -- in agreement with the XMM-Newton data, although these
were acquired one month later -- demonstrate variability mainly with
$\omega$. Due to the short 
duration of the observation -- slightly less than one orbit -- no secure
statement about the variability on the orbital period can be made. 

\subsection{Towards a refined spin ephemeris}
\label{section_timing}

Using the data acquired in 2005 and 2006, we attempt to improve the accuracy of the
spin ephemeris. The optical data of long-scale base line can be used to
refine the period, while the hard X-rays relate the ephemeris zero-point to
the rotation of the white dwarf.

From the spin-phase folded light curve of the R-band data from 2006
(Fig.~\ref{fig:rx0625_lc_08.05.2006}), a new time of spin maximum for a
long-term ephemeris can be derived. The phase of maximum, 0.075, corresponds
to $T_{\rm max, spin}$(BJED)$=2453864.5668(3)$. The error in the epoch is
assumed to be half the bin width, 0.025 spin phases. The cycle-count is lost
due to the error in the spin ephemeris from \citet{staude2003a}.

\citet{kim2005} proposed a refined spin ephemeris,
$P_{spin}$=$0.01374116815(17)$\,d, based on a 3-period fit. When folding our
R-band data from May 8, 2006 with their ephemeris, the spin maximum, however occurs at
phase 0.725(25), which is clearly outside the error ranges of their ephemeris
zero point at the time of observation (0.01 phases). The reason
for this discrepancy may be that they used data obtained in different accretion
states: this is obvious from the significant change in brightness between their measurements
in 2004 ($m_{\rm R}\sim14.5$) and 2005 ($m_{\rm R}\sim15.5$). We detected
fundamental differences in the variability patterns under similar
conditions, i.e.~a change from variation with frequency $\omega$ to frequency
$\omega$-$\Omega$. 
  
\begin{figure}[htb]
\includegraphics[height=\linewidth,angle=-90,clip]{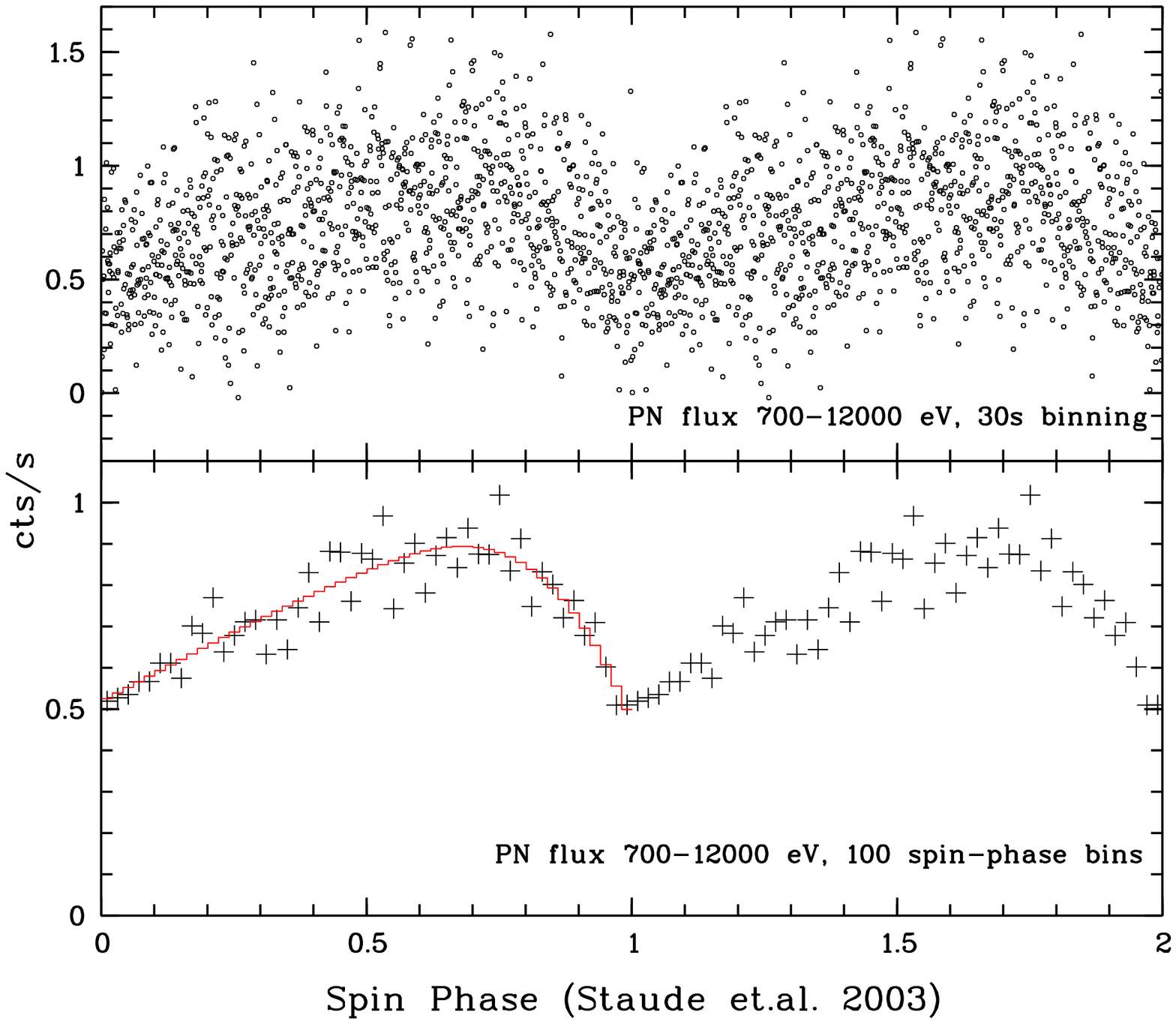}
\begin{minipage}{\linewidth}
\vspace*{0.5cm}
\end{minipage}
\includegraphics[height=\linewidth,angle=-90,clip]{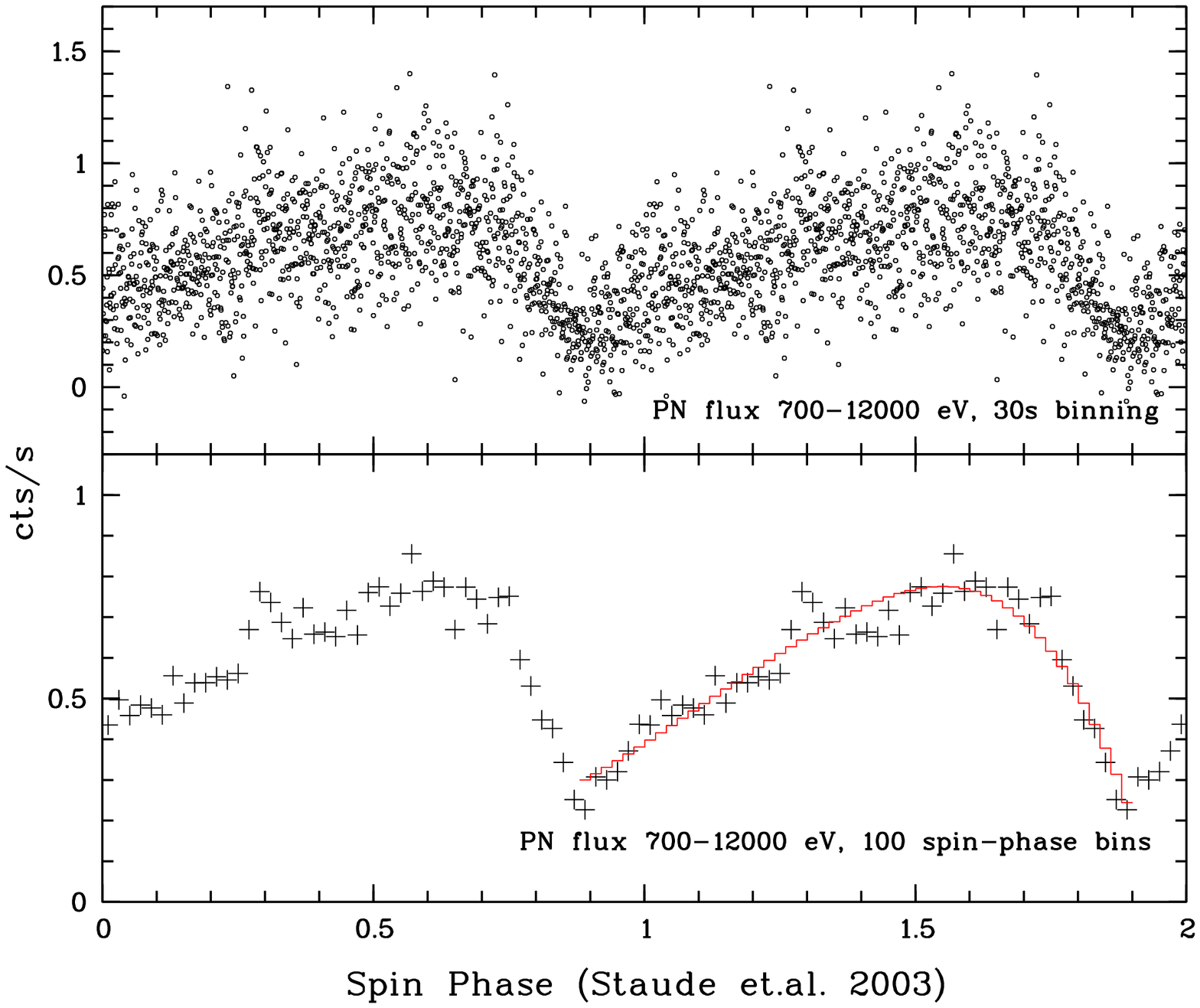}
\caption{\label{fig:PN_hard_spin_folded} The background-subtracted PN
  count rate in the energy band {\it MHX} (700 to 12000\,eV) ({\it top}
  2005/03/31, {\it bottom} 2006/04/06) binned to 30s, plotted over spin-phase
  ({\it upper}). The same data, binned into 50 phase bins with an overplotted
  spline fit ({\it lower}).} 
\end{figure}

We calculated an X-ray spin-ephemeris from phase-binned light curves of the
harder photons ({\it MHX}, $700 - 12000$\,eV,
  Fig.~\ref{fig:PN_hard_spin_folded}). Times of maximum flux were detemined
from a spline fit to the binned data to be \mbox{$T_{\rm max,PN_{hard}}($BJED$) =
  2\,453\,461.4475(8)$} (phase 0.68 $\pm$0.02) and \mbox{$T_{\rm
    max,PN_{hard}}($BJED$)  = 2\,453\,832.0615(8)$} (phase 0.57$\pm$0.02). The
assumed uncertainty in the maximum-determination is one bin-width. The flux
maxima were used because they represent the time of optimal visibility for the
principal accretion region of the white dwarf. Depth and position of the minimum may be
strongly dependent on the size of the accretion region(s), because their
extension determines the length and completeness of their occultation behind
the white-dwarf limb. The apparent optima visibility of the main accretion
region is only weakly dependent on the size.

The spin-ephemeris of \citet{staude2003a} provides a cycle-count of 26971,
which corresponds to a period of $P_{spin,MHX}$=0.01374120(6)\,d.  

This value matches the previously-published results well. 
The measurement of period confirms the previous estimate for an independent
data set, even though: (i) the period does not appear to be more accurate than
that derived from optical photometry in 2003; and (ii) the spin period
measured by \citet{kim2005}, is within the error range of our
results.  

The timing information derived from optical photometry cannot be combined with
that of the X-ray data, because the two types of radiation originate in
completely different parts of the binary system, even if their measured
variabilities have similar periods: a time-lag between the pulses is therefore
expected.

\subsection{From periodicity to structure}

In 2005, the spin-folded light curve in band {\it MHX} is sawtooth-shaped
(Fig.~\ref{fig:PN_hard_spin_folded}), with a slow rise to maximum in $\sim$0.7
phase units, while the decline lasts only 0.3 phase units. The light curves
have a pulsed fraction of 50 per cent and show a similar behaviour in the
energy sub-bands {\it MX} and {\it HX}.

In 2006, the {\it MHX} light curve looks similar to that from 2005, but its
minimum is more pronounced with a pulsed fraction of 75 per cent compared to
50 per cent in 2005. 

\begin{figure*}[htb]
\mbox{
\includegraphics[height=13.5cm,angle=0,clip]{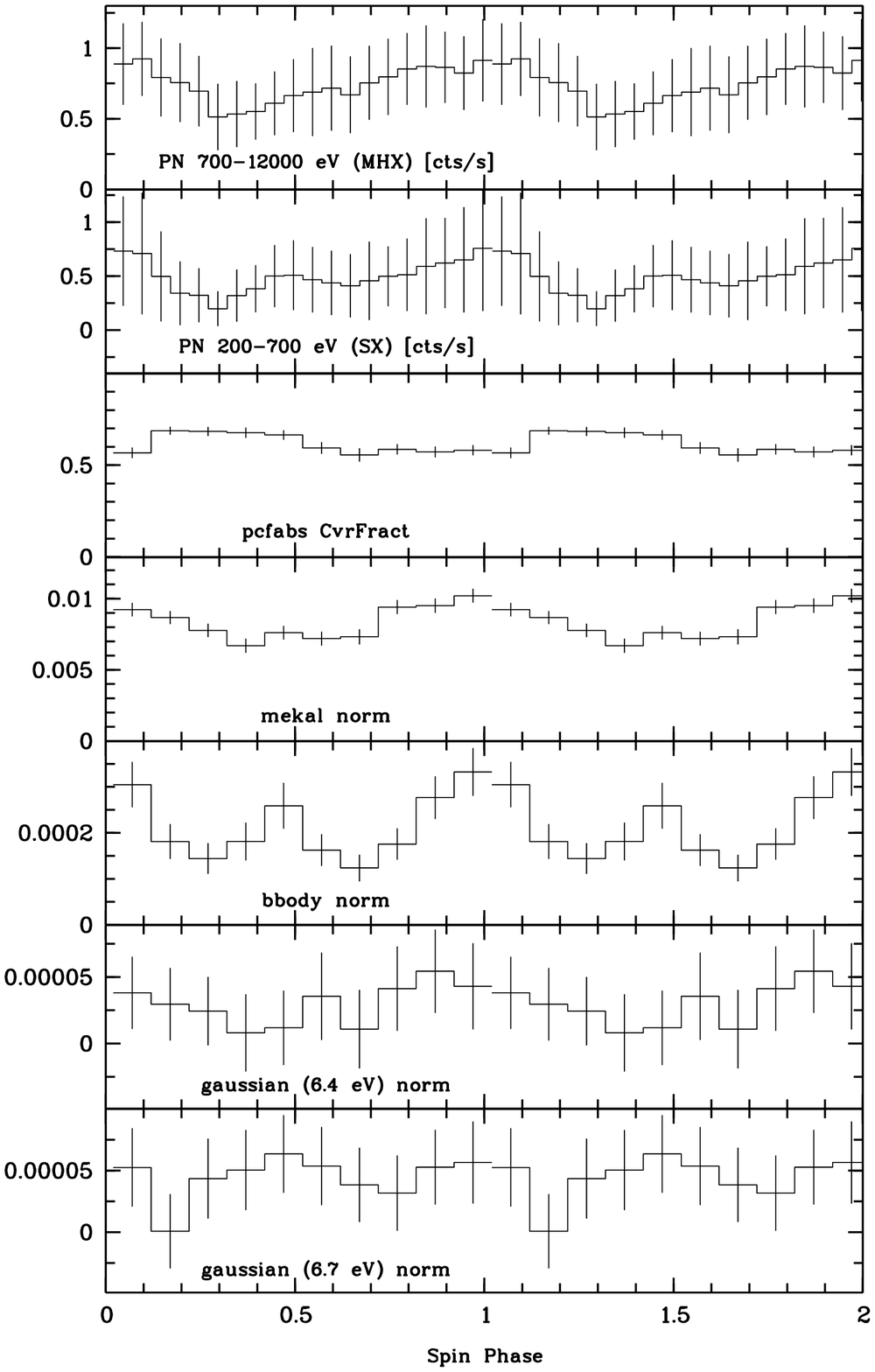}
\hfill
\includegraphics[height=13.5cm,angle=0,clip]{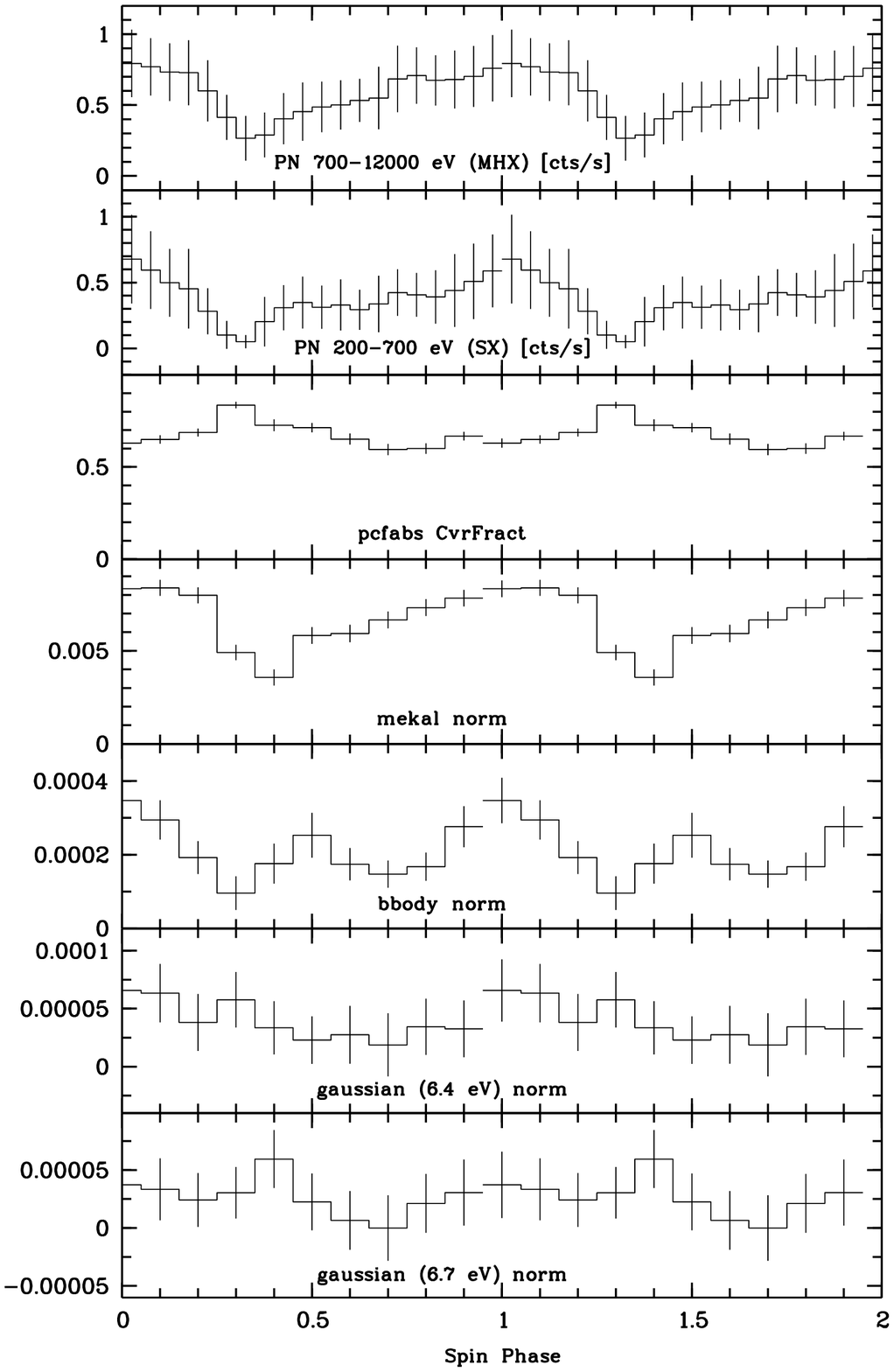}
}
\caption{\label{fig:PN_MOS_spinfolded} The spin-folded PN light curves 
  from for 700-12000\,eV ({\it MHX}) and 200-700\,eV ({\it SX}). For the
  calculation of the spin-phase, we use as zero points the epochs of maximum
  flux in the {\it MHX} band determined in Sect.~\ref{section_timing} and the
  period from \citet{staude2003a}. The error bars in the light curves mark the
  standard deviation of the values having entered the bin.  
Below are plotted the spin-phase dependent parameter values for the fit to
phase-selected MOS+PN spectra. The parameters not shown here were kept
constant at the values determined from the fit to the average spectrum. 
{\it left}: March 31, 2005, {\it right}: April 06, 2006
}
\end{figure*}

To obtain more information about the origin of the spectral components, a
phase-resolved spectral analysis was performed with respect to different
periodicities ($\omega$, $\Omega$, $\omega$-$\Omega$, $2\omega$-$\Omega$). The
data were binned into 10 phase bins for each period. A number of parameters
from the model fit to the global spectra were fixed to the values previously
obtained: {\it wabs\_nH} (since it is interpreted to be of interstellar
origin), {\it  pcfabs\_nH}, {\it mekal\_kT} (because this was, in any case,
not constrained), {\it 
  bbody\_kT} (which when allowed to vary, changed very little around its 
mean value), and {\it gaussian(1,2)\_sigma}.

The error bars in the binned light curve plots indicate the standard
deviation of values that enter the bin. We calculate the errors in our
phase-binned dataset by measuring the Poisson noise of photons in a given
bin. This is the correct approach when a system is observed repeatedly in a
similar state, which is clearly the case if data show variability only in the
period for which the data are binned. 

When binning data that are varied by multiple periods, the measurements
from different cycles, entering the same bin, are not in general obtained when
the system is in a similar state. If Poissonan noise was assumed to represent
the error, this would not take account of the intrinsic variability by the
superposition with the other periods, and would underestimate the errors.

In our data for MU Cam there is always variability on multiple periods -- in
the simplest case for just $\omega$ and $\Omega$. We could attempt to remove
the variation that occurs on one period from the data, but this will certainly
not produce the desired result in the presence of strong side-bands, as in the
{\it SX}-band in the 2005 data.

To maintain the size of errors in all plots of binned light curves, we
consistently use the standard deviation of values as they enter each bin
throughout the paper. These values represent upper limits to the true errors.
 
\subsubsection{The white dwarf rotation -- $\omega$}

The parameter values of the fits to the spin-phase selected spectra are shown
in Fig.~\ref{fig:PN_MOS_spinfolded} with phase-folded light curves in
the bands {\it SX} and {\it MHX} . The similar
shape of the soft phase-folded X-ray light curves is remarkable, despite the
obvious differences in the time domain (compare Figs.~\ref{fig:lc2005} and
\ref{fig:lc2006}). It appears that the strong variations in the 2005 data are
completely removed when the data are folded and averaged using the spin period. 

\subsubsection{The stars' orientation matters -- $\omega$-$\Omega$}

Variability at the beat-period $\omega$-$\Omega$ was detected only in the data
from 2005. It was the dominating period in the OM and the $R$-band light
curves, it was clearly present in the soft X-ray band ({\it SX}), whereas 
it was almost undetected in the hard X-ray bands.
The beat-phase folded X-ray light curves and
phase-resolved spectra do not show significant variation and are therefore not
shown.  

\subsubsection{A signature of two-pole accretion -- $2\omega$-$\Omega$}

\begin{figure}[htb]
\includegraphics[height=13.57cm,angle=0,clip]{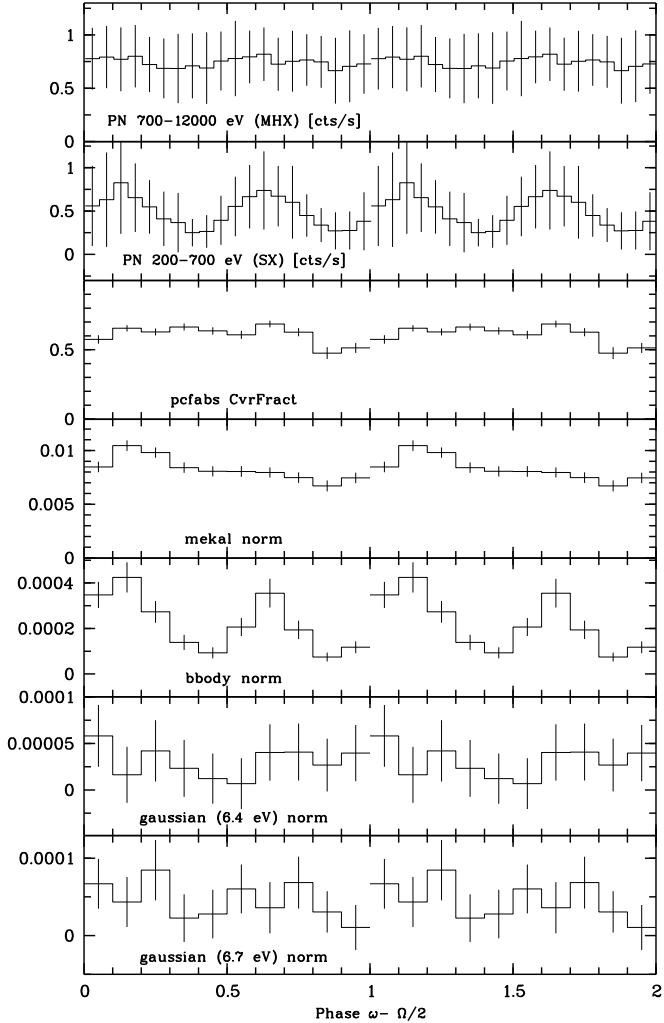}  
\caption{\label{fig:data2005_wynnfolded} Light curves in the hard and soft
  X-ray bands from 2005 folded on the period corresponding to 
    $\omega$-$\Omega/2$ (error bars in the light curves are the standard
  deviation of the values having entered the bin), and the free parameters of
  the fits to the phase-resolved spectra.  
}
\end{figure}

While the X-ray data from 2006 varies predominantly at the spin period,
the soft X-ray component in 2005 shows variation also with additional frequencies,
most prominently $2\omega$-$\Omega$. This is an indication that there are
two accreting poles, whose accretion rates are dependent on the orientation of
the white dwarf within the binary frame, i.e.~of both stars
\citep{wynnking1992}. To decide whether there is any difference between the
emission of the two accretion spots, the soft and hard X-ray photons were
phase-folded on the period corresponding to $\omega$-$\Omega/2$. Spectra were
extracted for ten phase-bins with respect to this period
($\Phi_{\omega-\Omega/2} = (BJED-2452682.4181) / 0.01423886$) and fitted. The
light curves and the fitted parameters are shown in
Fig.~\ref{fig:data2005_wynnfolded}.

Separated by 0.5 phases, there are two humps that have similar a soft X-ray
light curve, which explains the strong signal of frequency
$2\omega$-$\Omega$. The parameters 
of the fit indicate that the source of this variability are two distinct
poles: the {\it mekal} component has one peak only above an almost constant level,
and the integrated black body normalization of the humps clearly differ. 

\subsubsection{The orbital period -- $\Omega$}

\begin{figure}[htb]
\includegraphics[height=13.5cm,angle=0,clip]{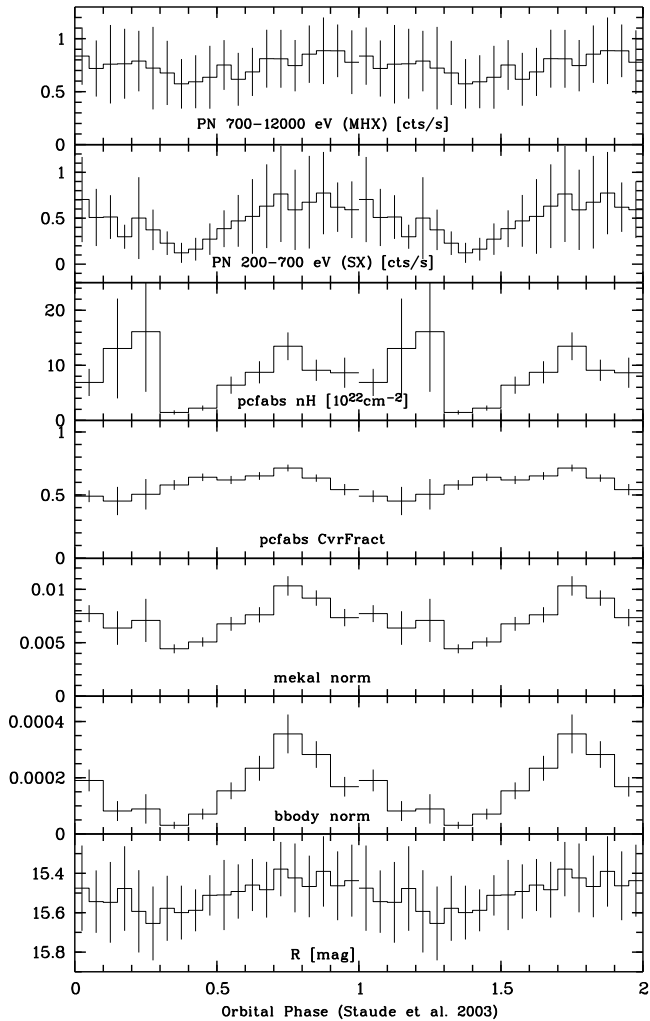}
\caption{\label{fig:data_orbfolded} The orbital period $\Omega$: Light
  curves from 2005 in the hard and soft X-ray bands, the fitted spectral
  parameters, and the phase-folded {\it R}-band data. 
The error bars are the standard deviation of the values having entered the
  bin.  
}
\end{figure}

For the data discussed here, a significant and likely periodic
dependency of the radiation on the orbital motion is detected  only for 
2005 (see Fig.~\ref{fig:aov}). Compared to the power in other detected
frequencies, the orbital signal is strongest in the soft X-rays, and 
detected clearly in the other X-ray bands and the optical. Phase-folded light
curves and fitted parameter values for the X-ray spectra from 2005, are shown
in the left panel of Fig.~\ref{fig:data_orbfolded}. There
is almost no variation in the UV light curve (which is therefore not shown),
but the other light curves display clearly a single hump. The maxima and
minima in the curves do not 
occur at the same orbital phases -- the R-band is 0.1 phases ahead. The probable
origin of the variability $\Omega$ is the changing visibility of emission sites
during the rotation of the binary. 

In contrast to the spectral fits applied to data averaged using the other
frequencies (see previous sections),
the parameter {\it pcfabs\_nH} was allowed to vary for studying the
$\Omega$-dependence of the data from 2005. 
By fitting the binned data, we searched for variations in the hydrogen column
density, caused by the changing orientation of the magnetically-guided stream
with respect to the line-of-sight during one revolution.
The number of observed orbital cycles, however is rather low for the binned data to 
be regarded as representative of the mean system; since fewer than two
consecutive orbits were observed, the data can be considered only as a
snap-shot of the system at a certain time.

The data from 2006, phase-folded on $\Omega$, are not shown, because only
marginal variability is observed in the X-ray light curves and OM data; 
the latter data set and optical data are affected by insufficient
coverage of the orbital cycle. We therefore do not attempt any phase-dependent
analysis.

\section{Discussion}

\subsection{Pure disk accretion observed in 2006}

In all of the X-ray periodograms, the dominant frequency is the white-dwarf spin
$\omega$. The frequency $2\omega$ is also present, but far weaker. In
the periodograms, there is signal at the orbital frequency $\Omega$ in all
wavelength ranges, although it is not prominent in X-rays and not
completely resolved in optical light.

Following the analysis of \citet{wynnking1992}, the natural explanation for
the detected variability pattern is accretion via an accretion disk and
accretion curtains, which connect the disk to the magnetic poles. The weak
detection of signal at frequency  $2\omega$, in addition to that at $\omega$, implies that
there are two accreting spots, almost opposite to each other 
Their properties or 
visibility must differ to explain the far stronger signal at
$\omega$. This interpretation is supported by Fig.~\ref{fig:PN_MOS_spinfolded}
(right panel), where the soft X-ray light curve shows (when compared to
the hard profile) a second maximum shortly before phase 0.5, and the normalisation
of the black body component of the phase-resolved spectral fits has a secondary
maximum at this position. The normalisation of the {\it mekal}-component shows only
one maximum -- located at the phase of maximum brightness -- with a slow rise
and a rapid decay. The sawtooth-like shape indicates its origin in an
arc-shaped region. The covering-fraction of the absorption component in the
spectral fits shows a smooth modulation on the spin period with a clear
maximum at the phase of minimum X-ray flux. 

The resemblance of the spin-folded X-ray light curves in the two energy ranges
to the ones shown by \citet{evans2004b} for V405 Aur is striking. As observed 
there, the soft X-rays show a double-hump, whereas the hard ones form a
sawtooth shape.
A similar accretion geometry, therefore, appears to exist in MU Cam, where the
white-dwarf dipole axis is located close to the orbital 
plane instead of being aligned with the spin axis.
The hard and soft X-ray light curves
and the parameters of the fits to the spin-resolved X-ray spectra
(Fig.~\ref{fig:PN_MOS_spinfolded}, right panel) can be explained in a 
model where the lower accreting pole has optimal visibility at spin phase
0.45, and the upper at phase 0. Between phase 0 and phase 0.3, the covering
fraction of the partial absorber increases. This absorbing matter is the
accretion curtain to the upper pole, crossing the line-of-sight to the X-ray
emitting region. The phase-shift between the maximum absorption and the optimal
visibility of the accretion region can be explained by the bulk of matter
being accreted along trailing field-lines, as proposed by
\citet{beardmore1998} to explain absorption features in observations of FO
Aqr. At phase 0.25, the shocked region above the upper pole, which emits
hard X-rays, disappears behind the white dwarf, in a similar way to the soft
X-ray spot. An indication that soft X-ray emission traces reprocession sites
of hard X-rays is the modulation of soft X-rays by the white-dwarf spin. 

In contrast to the optical data discussed in \citet{evans2004b}, our R-band data
from 2003 and 2006 do not show a double-peaked spin pulse. 
The spin-folded optical data from 2003 (see Fig.~4 in \citealt{staude2003a}) 
have a zigzag-like shape of amplitude $\sim$0.2\,mag. The spin-phase
folded R-band light curve from 2006 (Fig.~\ref{fig:rx0625_lc_08.05.2006})
shows a sinusoidal variation of amplitude 0.1\,mag. Using the hard X-ray
determined spin-ephemeris (Sect.~\ref{section_timing}), the flux maximum
for the 2006 data occurs at phase 0.55, i.e. half a spin cycle later than
the X-ray maximum. We propose that the pulsed fraction (10 per cent)
of the optical flux to be originates in the irradiated parts of the accretion
curtains. Their irradiated inner regions are most easily visible when the upper pole
points away from the observer. 
The magnetic axis of the white dwarf must be
tilted, with respect to the rotational axis, by approximately the system
inclination angle to be able to explain the maximum X-ray flux that is half a
spin-cycle offset from the optical maximum. When pointing towards the observer
at phase zero, this prevents obscuration of the upper accretion region by its
accretion curtain. 

Unfortunately, the OM light curve shows a clear variation pattern only towards
the end of the observation and is affected by data gaps; it is therefore 
impossible to cross-correlate the OM light curve with the X-ray and optical
light curves to measure the relative phasing, i.e.~the position of the
reprocessing sites.  

\subsection{Accretion via disk and stream observed in 2005}

The hard X-ray component is assumed to be originate in a shock inside the
accretion column above the magnetic pole. Since there is almost no variability
in the count-rate of the hard photons on the beat frequency $\omega$-$\Omega$,
the accretion rate does not appear to be dependent on the white-dwarf
orientation within the binary frame. The accretion producing the
emission of hard photons probably happens via an accretion disk, as occured in 
2006.

The number distribution of the soft X-ray spectrum ({\it SX}, $200 - 700$\,eV)
is reproduced well by a black body, and appear to
originate on the white dwarf surface (because of the strong signal
at $\omega$). The presence of $\omega$-$\Omega$ shows that the orientation
of the white dwarf in the binary frame has influence on the
emission. Stream-fed accretion, penetrating and heating the white-dwarf
surface, is probably therefore the origin of (parts of) the soft X-ray 
component, because pure reprocession of hard X-rays on the white dwarf would
produce a signal at $\omega$ only. The frequency $2\omega$-$\Omega$ implies
that there is a second accreting pole that is opposite to the principle pole,
since it denotes the detection of a similar accretion geometry after just a
little more than half a spin-cycle, in the case of stream-fed accretion
\citep{wynnking1992}. The two peaks per spin cycle can also be directly
observed in the light curve
(Fig.~\ref{fig:lc2005}). Figure~\ref{fig:data2005_wynnfolded}~shows that only
the soft X-ray flux (i.e.~the flux in the black-body component), not the hard
X-ray flux, is modulated with respect to the orientation of the two stars.  

As the dominant frequency in the optical and UV light in 2005,
the beat-frequency $\omega$-$\Omega$ indicates that the emission or visibility 
of this radiation component was strongly dependent on the orientation of the
white-dwarf magnetic field in the binary reference frame. 
The accretion disk may have been smaller in 2005 than in 2006, and the inner
rim therefore located at a greater distance from the white dwarf. 
This would allow a disk-overflowing part of the ballistic stream, which would
otherwise be negligible, to dominate accretion via the disk. 
Only the prolonged ballistic stream would have a sufficiently high density and
incident flux to re-emit significantly, producing a $\omega$-$\Omega$
periodicity. This view is supported by the lower brightness 
of the system in the optical/UV in 2005. It is observed that the amplitude of the
short-term variations did not vary significantly, in contrast to the size of
the constant component. 

The orbital variation in the X-ray emission (Fig.~\ref{fig:data_orbfolded}) is
probably caused by the different visibility of the accretion sites 
in the binaries frame of reference.
Our fitted parameters argue against a model, pushed forward by
\citet{hellier1991} to explain orbital modulation of the X-rays 
in AO Psc, where the minimum at orbital phase 0.3 is explained by increased
absorption, e.g.~by matter lifted out of the orbital plane.
Unfortunately, the published orbital ephemerides either have too
large errors \citep{sab2003,staude2003a} or a remaining cycle-count
ambiguity \citep{kim2005} to allow to relate the relative phasing of the
other detected periods to the probable epoch of inferior conjunction of the
secondary star, as determined by \citet{sab2003} from phase-resolved
spectroscopy. 

To explain the behaviour of the system during the XMM-Newton observation in
2005, we therefore propose that the white dwarf was accreting via an accretion
disk and a disk-overflowing stream. 
The hard X-ray emission, which is particularly sensitive to variations on the
white-dwarf spin period, are produced by shocks above the footlines of the
accretion curtains being fed by the accretion disk. The main (upper) accreting
pole is best visible around spin phase 0, when it points to the observer.  
The occurrence of the frequency $2\omega$-$\Omega$ in the soft X-ray band is
an indicator of two opposite, similarly bright, accretion spots, accreting via
an accretion stream. This is evidence of the white dwarf's dipole axis being
strongly inclined from the rotational axis, which produces  similar visibility
and accretion efficiency in both regions just offset by 0.5 phases in spin and
beat period, respectively. 

In the spectral model, we included two Gaussians to represent iron lines
at 6.4 and 6.7\,keV. Although they have no significant influence on the
overall $\chi^{2}$, they may support the interpretation of the spin-dependent
behaviour of the system. The normalisation of the \ion{Fe}{}-line at 6.7\,keV closely
follows the shape of the normalisation of the black body component with two
distinct peaks. The fluorescence line at 6.4\,keV appears to have only 
one peak and an eclipse-like feature at spin-phase 0.7, which 
resembles the {\it mekal} component. In general, this line is interpreted to be
caused by the reflection of the hard X-ray component by cold material, e.g. by the
irradiated surface of the white dwarf close to the accretion region. The coupling
of the normalisations of the {\it mekal} to the 6.4\,keV line supports this view in
the case of MU Cam. 

In the optical spectra published for MU Cam (\citealt{wei1999},
\citealt{sab2003}), there is no evience of spectral features that could be produced
on the secondary star. It is therefore impossible to estimate the contribution
of the late-type star to the overall flux and the distance of the system. Using
orbital period-spectral type relations for cataclysmic variables determined
either observationally \citep{smithdhillon1998} or theoretically
\citep{beuermannbaraffe1998}, the companion star in MU Cam is expected to be
of early M type (i.e.~$\sim$M1). 

\section{Summary and conclusions}

We have presented the analysis of comprehensive data sets obtained
simultaneously at optical, ultraviolet, soft and hard X-rays in 2005 and 2006
with the XMM-Newton observatory and the AIP 70cm telescope. Our main results
and conclusions may be summarized as following:

\begin{itemize}

\item The period tentatively derived as the white dwarf spin by
  \citet{staude2003a} is the dominant signal in the hard X-ray light
  curves. We claim that this provides a measure of the white-dwarf rotation. By
  linear interpolation of the timings of the mean hard X-ray pulse maxima in
  the two XMM-Newton observations we measure a spin ephemeris of $BJED(\rm
  max,PN_{hard}) = 2\,453\,461.4475(8) + E\times0.01374120(6)$ 

\item A distinct soft component in the X-ray emission, which was already
  visible in the ROSAT spectrum, was unambiguously detected and could be
  fitted well with a black body of temperature 59\,eV and 54\,eV in 2005 and
  2006, respectively. MU Cam could be reliably identified as a soft IP. 

\item MU Cam shows changes in its accretion state between disk-dominated and
  disk with an additional stream component during the observing interval from
  2003 to 2006. These are displayed by a change in the mean optical brightness
  of about 1\,mag, and by fundamental changes in the power spectra
  throughout the complete observed energy range. Such changes were also
  reported for other IPs (e.g.~TX Col: \citealt{norton1997}, FO Aqr:
  \citealt{beardmore1998}, \citealt{evans2004a}), and therefore appear to be a
  prevalent feature of intermediate polars.  

\item The occurrence of the $2\omega$ frequency in the X-ray data from 2006
  and the strength of $2\omega$-$\Omega$ in 2005 suggest two accretion regions
  at opposite positions with similar properties and visibility, at least in
  certain energy ranges. This may be achieved if the dipole axis of the
  magnetic field of the white dwarf was strongly inclined away from the
  rotation axis to the orbital plane. 

\item The strong modulation in all wavelength ranges implies a rather high
  inclination, whereas the absence of an X-ray (white dwarf) eclipse limits it
  to $\lesssim75\degs$. It is probably even lower, because no eclipse is
  detected in the optical, which would be a sign of a partial obscuration of
  the accretion disk. 

\item To improve our understanding of the properties of the system, future observations
  should include more optical photometry to refine the spin and orbital
  ephemerides. For studying the accretion geometry, time-resolved optical
  high-resolution spectroscopy will be crucial, and it may help to determine
  an accurate orbital ephemeris. To determine the distance,
  near-infrared spectroscopy should be performed to be able to determine the
  spectral type of the secondary star and its contribution to the spectral
  energy distribution. 

\end{itemize}

\begin{acknowledgements}

ASt acknowledges support by the Deutsche Forschungsgemeinschaft
under grant SCHW536/20-1, ASt, RSc, JV, and ANGM were supported by the DLR
under grant 50OR0404. 
We thank A.~Bird, C.~Knigge, and collaborators for
providing the {\it INTEGRAL} spectrum of MU Cam. 

\end{acknowledgements}

\bibliographystyle{aa} 
\bibliography{mucam_fin} 

\end{document}